\documentclass[aip,jcp,reprint]{revtex4-1}
\usepackage{mathptmx,amssymb}

\usepackage{amsmath}
\usepackage{graphicx}
\usepackage{comment}
\usepackage{color}

\let\oldhat\hat

\renewcommand{\hat}[1]{\oldhat{\mathbf{#1}}}

\begin{document}

\author{Vitaliy Ogarko, Nicolas Rivas, and Stefan Luding}

\affiliation{Multi Scale Mechanics (MSM), MESA+, CTW, University of Twente, PO Box 217, 7500 AE Enschede, The Netherlands.}

\title{Structure characterization of hard sphere packings in amorphous and crystalline states}


\begin{abstract}
The channel size distribution in hard sphere systems,
based on the local neighbor correlation of four particle positions,
is investigated for all volume fractions up to jamming.
For each particle, all three particle combinations of neighbors define channels, which are relevant for the concept of caging.
The analysis of the channel size distribution is shown to be very useful in distinguishing between gaseous, liquid, partially and fully crystallized,
and glassy (random) jammed states.
A common microstructural feature of four coplanar particles is observed in crystalline and glassy jammed states, suggesting the presence of ``hidden'' two-dimensional order in three-dimensional random close packings.
\end{abstract}

\maketitle

\section{Introduction}

%
%


The hard-sphere particle interaction limit is a tremendously versatile physical model, being widely used for structural studies of liquids\cite{hansen_theory_2006}, glasses \cite{alder_phase_1957,bernal_geometry_1960}, colloids \cite{russel_colloidal_1992}, granular materials \cite{duran_sands_2000}, and many others \cite{kalos_helium_1974,torquato_effective_1986}.
Its relevance in such a variety of physical systems suggests that many macroscopic properties arise by the fundamental fact of impenetrability of the systems constituents \cite{chaikin_principles_2000}.
The ultimate goal then becomes to establish relations between physical properties and the geometry of the arrangement of hard bodies in two or three-dimensions.
Many decades of research, heavily driven by numerical experiments\cite{alder_phase_1957},
have led to various geometrical structure variables, with different levels of success in either uniquely characterizing each state
or in deriving macroscopic physical properties from them \cite{TorquatoBook}.

Our study is motivated by the structural phase transitions observed in molecular fluids
and also replicated in hard-sphere systems under compression \cite{Rintoul1996}.
We also consider analogous phenomena observed in granular materials,
where the hard-sphere approximation is commonly used 
to successfully model complex rheological behaviours \cite{duran_sands_2000}.
As the volume fraction is increased, hard-spheres enter an entropy minimization driven phase
where glass formation competes with the nucleation and growth of the crystalline phase \cite{omalley_crystal_2003}.
Hard-sphere models are known to successfully reproduce the main structural properties of these states for various physical systems,
either for crystallization \cite{hoover_melting_1968, auer_numerical_2004},
or the amorphous solid phase transition \cite{rintoul_hard-sphere_1998}.
One of the main reasons for using the hard-sphere model over classical condensed matter systems
is that simulations -- as also colloidal suspensions experiments, which are a very good approximations of the hard-sphere model \cite{auer_numerical_2004, Pusey2009}
-- have particle-size spatial resolution, and thus the statics and dynamics can be studied from a microscopic perspective.
Furthermore, due to the lack of long-range or non-binary interactions, and the simple geometry of the constituents, hard-sphere models are theoretically tractable \cite{wertheim_exact_1963, hoover_melting_1968, deHaro2008}.

Inspired by the highly ordered and easily describable crystalline phase, many researchers have searched for inherent geometrical relations in disordered (amorphous) packings.
The straightforward approach is to analyze the static structure factor, which is a direct measure of the local microstructure of particles \cite{Cusak1987}.
The pair correlation function $g(r)$ is also a popular quantity for the analysis of non-crystalline materials \cite{Donev2005PRE, Estrada2011}.
The problem with such quantities is that the detailed three-dimensional information is lost as a result of statistical averages, as also by considering only pairs of particles.
In particular, they do not provide much information about the topology of the local structures in the particle-size scale,
which are believed to distinguish different kinds of amorphous arrangements.
It therefore becomes highly significant to exploit some other methods of three-dimensional characterization of these structures.
Many attempts have already been made to quantify local or long-range ordering, providing further characterization of disordered packings
\cite{Finney1970, Finney1981, Bywater1983, Hiwatari1984, Kimura1984, Chan1990, Voloshin2002, Philipse2003, Haw2006, Anikeenko2007, anikeenko_shapes_2008}.


In the following we analyze local arrangements of particles recognizing the importance of caging and voids in the overall structure and properties of the arrangement.
Here we extend the previous studies \cite{Finney1981, Bywater1983, Chan1990} by considering all particle triples in the particles neighbourhood that do not include the central particle.
The distribution of voids allows us to clearly distinguish between the different structural phases,
as also between different kinds of crystals and the relative number of each specific ordering, in systems presenting partial or many types of crystallization.
Our analysis shows that in amorphous states there is a preferred local structure of four co-planar particles.

\section{Method of analysis}
\label{method}

We consider systems of $N$ non-overlapping spheres arranged in a three-dimensional cubic space of volume $V$.
The spheres are located at positions $\mathbf{x}_{i}$ and have radii $r$.
Our main parameter is the sphere volume fraction, defined as $\nu = (4/3)\pi \sum_{1}^{N} r^3 / V$.

\begin{figure}
\begin{center}
\includegraphics[width=0.98\columnwidth]{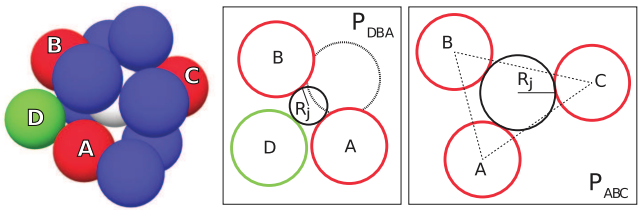}
\caption{(left) The central particle (white) is shown together with its nearest neighbours, defined by Delaunay-edges.
The channels for neighbour-triangles DBA (middle) and ABC (right) are shown in the neighbor-triangle plane.
The particles A, B and C are lying almost on the same plane with the central particle and are practically touching it,
so the channel almost coincides with the central particle. This is not the case for the BDA triangle.
}
\label{fig:circles}
\end{center}
\end{figure}

To determine the neighbors of a particle, we first compute the weighted Delaunay triangulation
of the set of points corresponding to the centers of the particles, $\{ \mathbf{x}_{i} \}$ \cite{PourninThesis}.
Neighboring particles are then defined as those particles connected by the edges of the triangulation.
For each particle we consider every possible combination of three neighbors, that is,
all possible triangles that can be formed by the centers of any three of its neighbors.
We refer to these triangles as neighbor-triangles.
Notice that the neighbor-triangles do not contain the central particle.
We then proceed to quantify the overall mobility of the particle by defining all channels through which the particle can pass.
A channel is defined as the circle-area in the plane of a neighbor-triangle through which the central particle could move.
This is computed by considering the Apollonius circle, i.e, the circle which is simultaneously tangent
to all other three circles defined by the projection of the three spheres in the neighbor-triangle plane.
There are at most eight possible Apollonius circles for each case, which are obtained analytically by solving a system of three quadratic equations \cite{Courant1996}.
From the set of eight possible solutions we choose the one which corresponds to the circle that does not contain any particle center of the neighbor-triangle,
as it is the only one that corresponds to our definition of channel
\footnote[1]{It may happen that no such circle exists, which was occasionally observed for polydisperse systems.
In this case we skip the corresponding neighbor-triangle from the statistics.}.
The radius of the respective channel is then defined as the radius of this circle, $R_j$, 
as shown in Fig.~\ref{fig:circles}.

Having obtained $R_j$ for all neighbor-triangles of every particle,
we then compute the normalized probability distribution function of (scaled) channel sizes, $f(R_j/r)$.
The ratio $R_j/r$ is calculated for all neighbor triples $j$ with channel size $R_j$ of every particle.
Note that $R_j/r$ has a direct physical interpretation, as less than unity corresponds to a closed channel, while greater than unity corresponds to an open channel, through which eventually the central particle could escape.
Furthermore, the function $f$ is well defined for spheres with any size distribution, since the radius of the central particle is scaled out.
We analyze both the individual structure of $f$ as also its evolution with volume fraction, for various particle systems.

\begin{figure}
\begin{center}
\includegraphics[width=0.8\columnwidth]{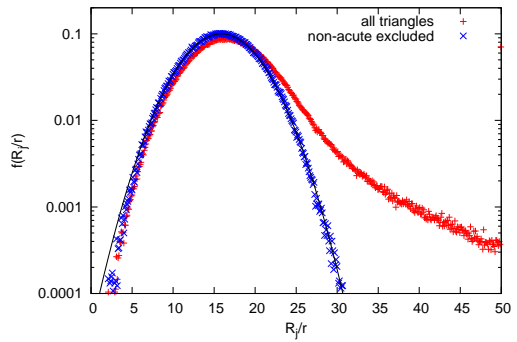}
\caption{Normalized distribution of the channel sizes scaled with the particle radius in the gas regime ($\nu \approx 0.0014$)
using full statistics (red pluses) and with non-acute neighbor-triangles excluded (blue crosses).
The solid line is a Gaussian fit $g(x)=(\sigma \sqrt{2\pi})^{-1} \exp \left [-(x-\mu)^2/(2\sigma^2) \right ]$
with parameters $\sigma \approx 4$ and $\mu \approx 15.9$.
The bin-size is 0.1.}
\label{fig:gas}
\end{center}
\end{figure}

In order to refine our definition of channels, we consider $f(R_j/r)$ for very low volume fractions, where no structure is expected (see Fig.~\ref{fig:gas}).
When considering all triangles the distribution presents a recognizable wide tail structure,
but after excluding from the distribution the channel sizes that correspond to non-acute neighbor-triangles,
i.e., those where one of the angles is greater than 90 degrees,
the distribution becomes Gaussian, with high accuracy over three orders of magnitude
\footnote[2]{We use the following test to check if a triangle with edge lengths $a$, $b$ and $c$ is acute:
$(a^2 + b^2 > \epsilon c^2) \& (c^2 + b^2 > \epsilon a^2) \& (a^2 + c^2 > \epsilon b^2)$,
where $\epsilon = 1-10^{-12}$ is used to account for numerical error.}.
The exclusion of non-acute triangles makes physical sense considering that channels defined by them cannot block the central particle,
thus conflicting with our initial definition of a channel.
For the rest of the analysis, non-acute triangles are never considered.

\section{Simulation details}
\label{simul}

We use an event-driven molecular dynamics algorithm, as it is fundamentally suited for the simulation of hard spheres systems.
The number of particles is by default $N = 16^3 = 4096$, unless stated otherwise.
Given the large amount of possible neighbor-triangles for each particle,
the statistical significance rapidly increases with the number of particles in the system.
We observed that $4096$ particles was adequate, as increasing the number of particles did not produce any noticeable change in any of the results.
Periodic boundary conditions are imposed to mimick an infinite system, i.e., a statistically homogeneous medium.

Starting from zero volume fraction, we compress the system towards a jammed state using a modification of the Lubachevsky-Stillinger algorithm \cite{lubachevsky90, donev2005},
which allows the radius of the particles to grow linearly in time with a dimensionless rate $\Gamma$
\footnote[3]{The growth rate is defined as $\Gamma = \frac{da}{dt} \sqrt{\frac{3M}{2E}}$ with the total system mass $M$.}. The kinetic energy, $E$, is kept constant using a re-scaling thermostat procedure \cite{Ogarko2012, Ogarko2013}.

If the growing is sufficiently slow, $\Gamma < 0.0007$ \cite{Hopkins2012},
the monodisperse system stays in a gas-fluid state in approximate equilibrium during the densification phase,
and exhibits a fluid-solid transition (crystallization) for volume fractions between $\nu_{\text{f}} \approx 0.492$ (freezing point)
and $\nu_{\text{m}} \approx 0.543$ (crystal melting point).
For infinitely slow compressions, it is expected that the system finally reaches a stable solid (crystalline) phase with close-packing fraction $\nu_{\text{cp}} \approx 0.7405$, corresponding to face-centered close packing.
In our simulations, due to finite compression rates, we reach a crystalline phase with defects and different local arrangements, and packing fractions up to $\nu \approx 0.73$.
This corresponds to a thermodynamically stable branch in the hard sphere phase diagram \cite{Rintoul1996}.
On the other hand, for fast compression rates the system enters a \textit{metastable} state for $\nu > \nu_{\text{m}}$,
which extrapolates continuously from the fluid branch and is conjectured to end at some random close packing state, around $\nu_{\text{rcp}} \approx 0.64$,
the interpretation of which is beyond the scope of this study, as its value depends on the details of the procedure \cite{Kumar2014}.

\section{Results and discussion}

We now observe the evolution of $f(R_j/r)$ with $\nu$ for fast and slow compression rates.

\begin{figure}
\begin{center}
\includegraphics[width=0.8\columnwidth]{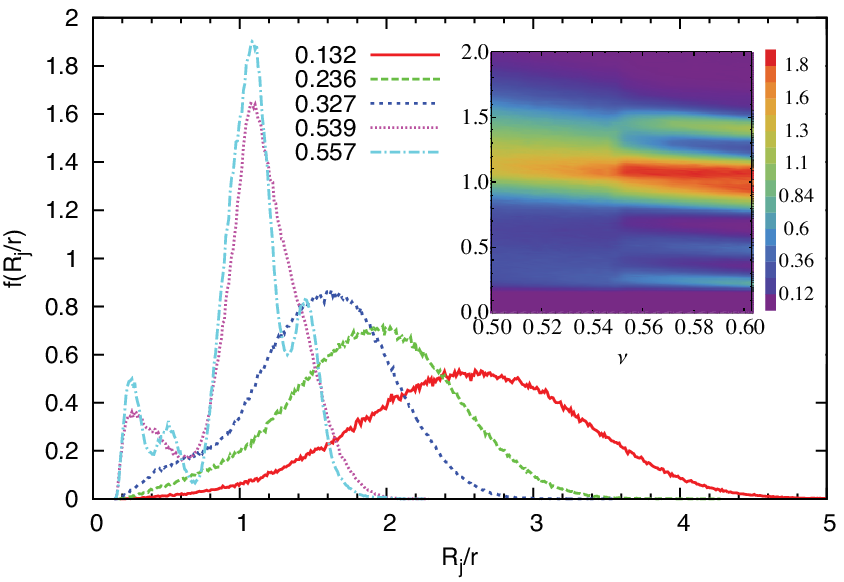}
\includegraphics[width=0.8\columnwidth]{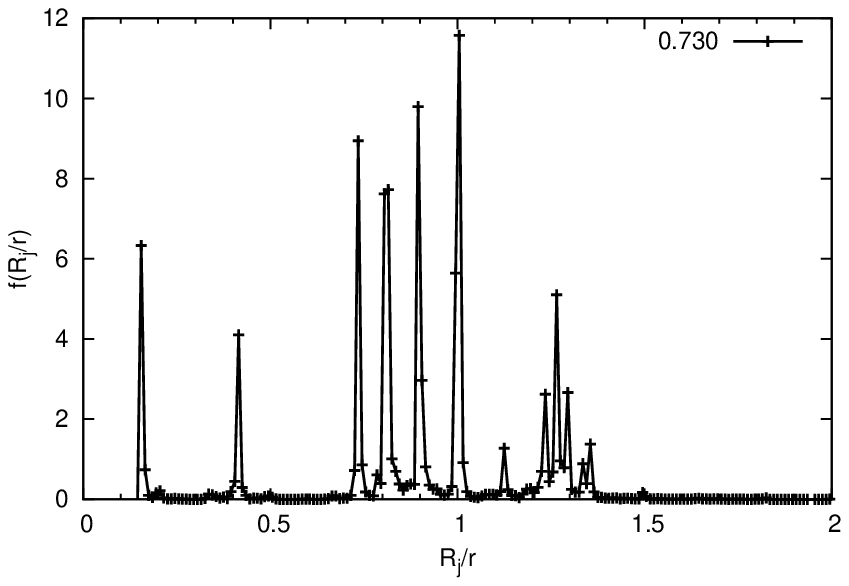}
\caption{Normalized distribution of the channel sizes scaled with the particle radius for slow compression ($\Gamma = 16 \times 10^{-6}$) and various volume fractions given in the inset. 
The bin-size is 0.01.}
\label{fig:xtal}
\end{center}
\end{figure}

\subsection{Crystallization path}

As the volume fraction increases, the distribution of channel radii fundamentally changes, see Fig.~\ref{fig:xtal} (a).
The distribution changes to non-Gaussian for fluid densities above $\nu \approx 0.15-0.25$.
We speculate that this change corresponds to the percolation gas-to-fluid transition observed by Woodcock \cite{Woodcock2012} at similar packing fractions,
although we did not investigate this in detail.
As the volume fraction increases, two smooth humps continuously grow,
that at higher $\nu > 0.5$ evolves into two well defined peaks, centered above $R_j/r \approx 0.15$ and near $R_j/r = 1$.
These values can be understood in terms of the geometry of the local arrangements:
$R_j/r \approx 0.15$ ideally corresponds to the channel size expected for three touching equal spheres,
and thus the appearance and growth of this peak shows the appearance of triples in contact as well as the relative importance of density fluctuations.
It is also the absolutely smallest possible channel size for equally sized spheres.
The peak at unity, on the other hand, is obtained for three particles
lying on the same plane with the central particle and practically touching it, i.e., when the channel essentially coincides with the central particle;
we confirmed that the majority of particles corresponding to the peak at unity are indeed practically touching the central particle.

As expected for very slow compression, $\Gamma = 16 \times 10^{-6}$,
the system exhibits (partial) crystallization near the melting point $\nu_{m} \approx 0.54$;
crystallization at the freezing point is kinetically suppressed \cite{Donev2007}.
The distribution is able to capture the crystallization transition by the development of two new peaks, at $R_j/r \approx 0.4$ and $R_j/r \approx 1.4$.
This transition is shown with more detail in the inset of Fig.\ \ref{fig:xtal} (a).
The new peak at $R_j/r \approx 0.4$ corresponds to a square crystalline arrangement.
On the other hand, the peak at $R_j/r \approx 1.4$ groups several distinct arrangements which
can be distinguished as $\nu$ further increases, as shown below.
This was confirmed for many runs with different initial particle velocities and positions.

\begin{figure}
\begin{center}
\includegraphics[width=0.48\columnwidth]{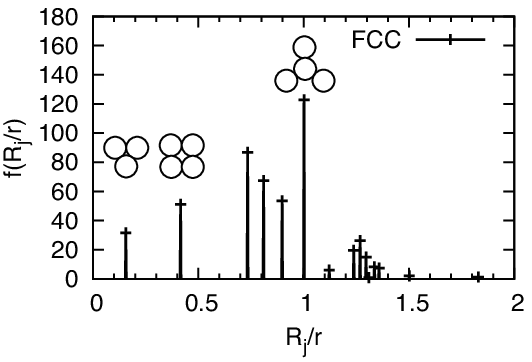}
\includegraphics[width=0.48\columnwidth]{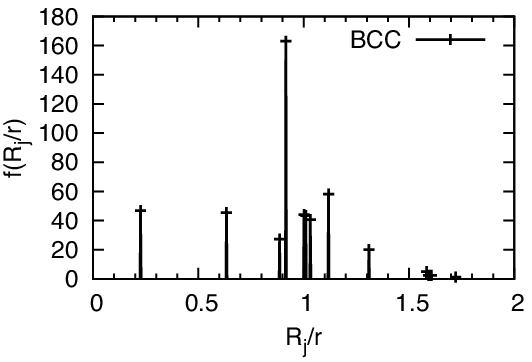}
\\
\includegraphics[width=0.48\columnwidth]{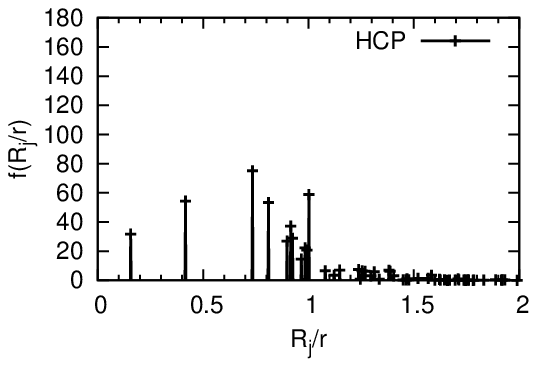}
\includegraphics[width=0.48\columnwidth]{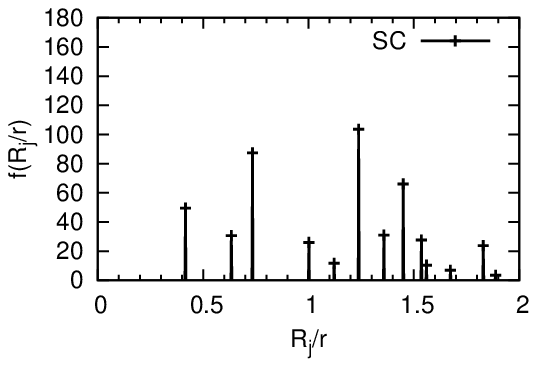}
\caption{Normalized distribution of the channel sizes scaled with the particle radius for perfect FCC, BCC, HCP and SC crystals with
1099, 1024, 1254 and 1000 particles, respectively. The bin-size is 0.002.
 Insets in the FCC case show triangle, square, and the typical local ordered described in the next, projected in the plane, at their corresponding peaks.}
\label{fig:xtals}
\end{center}
\end{figure}

Finally, as the maximum volume fraction is reached (i.e, as the pressure diverges),
the distribution is mostly dominated by steep peaks, see Fig.\ \ref{fig:xtal} (b).
In analogy with Bragg peaks from common diffraction techniques, 
these peaks can be traced to the crystal structures present in the particles' arrangement.
Figure \ref{fig:xtals} shows $f(R_j/r)$ for perfect FCC, HCP, BCC and SC crystals of about $10^3$ particles, as a reference. They all present significant differences, which allows to distinguish between the types of crystals. A detailed discussion of these peaks is beyond the scope of this study.

\begin{figure}
\centering
\includegraphics[width=0.8\columnwidth]{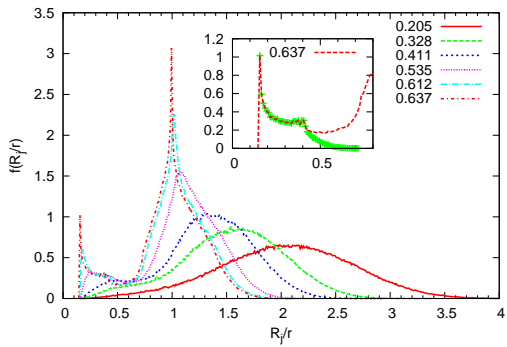}
\caption{Normalized distribution of the channel sizes scaled with the particle radius for fast compression ($\Gamma = 16 \times 10^{-3}$) of a monodisperse system with various volume fractions $\nu$ given in the inset.
The region of low $R_j/r$ is zoomed in the inset for the curve with $\nu = 0.637$.
The data with non-Delaunay neighbour-triangles excluded for the system with $\nu = 0.637$ are shown in the inset with green pluses.
The bin-size is 0.01.}
\label{fig:glass}
\end{figure}

\subsection{Glassification path}

Let us now focus on the case of fast compression, $\Gamma = 16 \times 10^{-3}$,
for which crystallization is not happening.
The channel size distributions for different volume fractions are shown in Figure \ref{fig:glass}.
For volume fractions below $\nu \approx 0.2$, there is no appreciable difference with the slow compression case described in the previous subsection.
For values well above $\nu \approx 0.2$, the same peaks at $R_j/r \approx 0.15$ and unity can be observed, but no other peaks are developed.
Contrary to the crystallisation path at the melting point, $f(R_j/r)$ does not show sharp signs of the glass transition at $\nu \approx 0.57$.
The lack of defined peaks in the distribution, as in the previous slow compression cases, is a clear signal that the system remains, to very high degree, amorphous.
Nevertheless, there exists partial order, as suggested by the high values of the peak at unity.
While the peak at $0.15$ can be interpreted in the same way as in the slow compression rate case,
the peak at unity, on the other hand, is not as easily interpretable.
While present in both crystalline and glassy configurations, its relative importance and shape are considerably different.
The unity value corresponds to those configurations where three neighboring particles
all touch the central particle and lie in the same plane with it.
There are of course other cases where the value of $R_j/r$ could be one.
By looking at the distribution of distance of each neighbor-triangle particle to the center particle, we confirm that by far the most common case is when the three particles are indeed touching the central one.
This suggests the existence of ``hidden'' local order in random close packings
that cannot be easily measured by order parameters because such local planes,
corresponding to $R_j/r \approx 1$,
are not oriented with respect to each other as in a periodic crystal structure.
We have no explanation for this preference of the system structure, suggesting a direction of future research. Furthermore, we have confirmed that the same qualitative features are observed in polydisperse systems\cite{VitaliyThesis}.

Let us now take a look at the structure of $f$ for values close to the minimal channel size, shown in the inset of Figure \ref{fig:glass}.
The high values between the two ideal cases, corresponding to the three touching spheres ($R_j/r \approx 0.15$) and
a square arrangement ($R_j/r \approx 0.41$), signals a significant presence of ``intermediate'' configurations.
The drop at $R_j/r \approx 0.41$ is due to exclusion of non-acute neighbor-triangles from the statistics.
A similar distribution of channel sizes, up to $R_j/r \approx 0.41$, was obtained
in the studies of interstitial holes in random close packings of spheres \cite{Finney1981, Bywater1983, Chan1990}. They concluded that for monodisperse packings the spread in the channel sizes between $0.15$ and $0.41$ cannot be reduced to lead to a single distribution of channel sizes allied with mechanical stability.

\section{Conclusions}
\label{conclusions}

The analysis of channel size distributions was shown to be able to distinguishing between gaseous, liquid, partially and fully crystallized, and glassy (random) jammed states.
Unlike the usually computed pair-distribution functions or structure factors,
the channel size distribution is highly sensitive to changes in volume fraction, 
and presents unique features for each phase.
States of partial crystallization can be recognized and characterized by the development and position of specific peaks,
which can be traced to specific crystalline configurations, and could be used to quantify the degree of crystallization of the system.
On the other hand, we confirm that random glassy configurations of isotropically (rapidly) grown hard particle systems present a common structural feature,
as shown by looking at the channel size distributions.
The overpopulation of many three neighboring particles lying in the same plane as the central particle, almost touching it, could be considered a first microscopic trace of crystals in a plane.
As different planes are not oriented with dominant relative angles as in a crystal,
there is no appreciable global three dimensional ordering. Lastly, we remark that our analysis
should be easily extendable to other particle systems even in different dynamic regimes.
Further work on the behaviour of the distribution near the state transitions could lead to deeper relations with previously known distribution functions or thermodynamic variables.

\section*{Acknowledgements}

We would like to thank J.\ Finney and L.\ Woodcock for helpful discussions.
This research is supported by the Dutch Technology Foundation STW, 
which is the applied science division of NWO, and the Technology
Programme of the Ministry of Economic Affairs, project number STW-MUST 10120.


\bibliographystyle{ieeetr}

\bibliography{bibliography}

%
%

\end{document}